\documentclass{ws-procs975x65}

\begin{document}

\title{Multiwavelength Observation of WIMP Annihilation}

\author{Roberto Aloisio}

\address{INFN - Laboratori Nazionali Gran Sasso \\
SS 17 bis, Assergi (AQ) Italy, \\ 
E-mail: roberto.aloisio@lngs.infn.it}

\maketitle

\abstracts{The annihilation of neutralino dark matter may result in 
observable signals in different wavelength. In the present paper we 
will discuss the effect of neutralino annihilation in the halo of our
Galaxy and in its center. According to high resolution cold dark matter 
simulations, large virialized halos are formed through the constant 
merging of smaller halos appeared at previous times. At each epoch, dark 
matter halos have then a clumpy component which is made of these merging 
subhalos. The annihilation of dark matter in these clumps, always present in 
the halo of our Galaxy, may be responsible for appreciable fluxes of 
$\gamma$-rays, potentially detectable. We find that, depending on the 
fundamental parameters of the clump density profile and on the distribution
of clumps in the Galactic halo, the contribution to the diffuse $\gamma$-ray 
background from clumps could be used to obtain constraints on the neutralino 
properties such as mass and annihilation cross section. On the other hand
the annihilation of neutralino dark matter in the galactic center may result 
in radio signals. At the galactic center, infact, the accretion flow onto the 
central black hole sustains strong magnetic fields that can induce synchrotron 
emission, in the radio wavelength, by electrons and positrons generated in 
neutralino annihilations during advection onto the black hole. We find that 
the observed emission from the galactic center is consistent with neutralinos 
following a Navarro Frenk and White density profile at the galactic center 
while it is inconsistent with the presence of a spike density profile, supposed
to be generated by the formation history of the central black hole.
}

\section{Introduction}

Most of the matter in the universe has yet to be observed in any
frequency band, thus the name, dark matter (DM). The evidence for the 
predominance of  dark over visible matter comes mainly from the
gravitational effects of the dark  matter component. However,
gravitational studies have been unable to shed light on the nature of
the dark matter. Big bang nucleosynthesis constrains  most of the dark
matter to be of non-baryonic origin. This has encouraged the study of
plausible new particle candidates for the dark matter.

Weakly interacting massive particles (WIMPs) are natural candidates for the 
dark matter. Particles with masses around $\sim 100$ GeV that interact only 
weakly have freeze-out densities in the required range of densities. 
In addition, particle physics models that invoke supersymmetry generate a 
number of plausible WIMPs. In the supersymmetric extensions of the standard 
model, the lightest supersymmetric particle may be stable due to conservation 
of R-parity enabling their survival to the present. In addition to massive and 
weakly interacting, dark matter particles are expected to be neutral. A class 
of neutral lightest supersymmetric particles is represented by a combination 
of gauginos and higgsinos, named the neutralino often represented by $\chi$.

Given the requirements of neutralino production in the early universe, it is 
possible to study the phenomenology of such dark matter candidates in detail 
\cite{jun}. In particular, the annihilation of neutralinos has often been 
considered a potential source of detectable secondaries: high energy particles 
and electromagnetic radiation. In this sense, dark matter can be visible 
through the radiation caused by the annihilation secondaries 
\cite{ber1,ber2,gon1,berg1,gon2,cal,bert,berg2,pas1,tyl,ull,oli1,noi1,ces,noi2}. 
Since the neutralino is a Majorana particle, it will 
self-annihilate at a rate proportional to the square of neutralino density. 
Thus, as was first realized in \cite{ber1}, the highest density dark matter 
regions are the best candidates for indirect searches. In the present paper 
we will review the annihilation signal that could come from the clumped halo 
component of our Galaxy and from its Galactic Center (GC), in these regions 
infact it is expected a large enhancement in the neutralino density with a 
consequent amplification of the annihilation signal.

Recent advances in Cold Dark Matter (CDM) simulations have shown that the
large scale structure of the Universe can be explained in terms of a 
hierarchical scenario in which large halos of dark matter are generated by 
the continuous merging of smaller halos \cite{ghi,moo,kly}. In this picture
a dark halo is the superposition of a smooth component, characterized by a 
typical scale comparable with the virial radius of the forming structure, 
and a clumped structure made of thousand of small scale halos. 
CDM simulations also show that most halos are well 
described by a density distribution with cusps at the center of each
halo.  The exact shape of the central cusp is still a matter of debate.
Most recent simulations favor profiles with density cusps varying from
the  Moore et al. profile \cite{moo} where $\rho_{DM} (r \to 0)\sim
r^{-1.5}$ to a Navarro, Frenk, and White (NFW) profile
\cite{nfw} where $\rho_{DM} (r \to 0) \sim r^{-1}$.

The debate is exacerbated by 
observations of galaxy rotation curves that seem to provide no evidence 
of central cusps \cite{sal}. However, the survival of 
cusps in galactic centers is highly dependent on the galaxy's merger 
history in particular on the history of formation of galactic center 
black holes \cite{mer}.
The central regions of small mass dark matter halos (DM clumps)
are less affected by the dynamics of baryonic matter and less likely to
have black hole (BH) mergers at their centers. Thus, a cuspy profile may well
describe the density of DM clumps. 

Another important piece of information is embedded in the spatial distribution
and survival history of clumps on their way to the central part of the 
host galactic halo. Much physics is involved in the description of the
structure of a DM clump moving in a larger DM halo, and different 
recipes are possible. These unknowns have forced us to consider two 
different scenarios, that for simplicity we call type I and type II
scenario.

In the type I scenario, the clump position in the host Galaxy determines its 
external radius. In the case of NFW and Moore profiles, the location of the
core is then determined by assuming a fixed fraction of this external radius.
In the type II scenario, a recipe is taken from the literature for the 
so-called concentration parameter, defined as the ratio of the core and 
virial radii of a clump. The recipe allows one to determine the properties 
of a clump for a given mass. We show that in this scenario the clumps are 
gradually destroyed on their way to the center of the Galaxy, so that an 
inner part, depleted of its clumpy structure is formed. This seems compatible 
with numerical simulations \cite{pow}. The effects of 
these two scenarios on the $\gamma$-ray emission from the annihilation of 
CDM particles are dramatic: in the type I scenario, the highly concentrated 
clumps that are implied produce strong $\gamma$-ray emission while in the 
type II scenario the low concentration clumps imply $\gamma$-ray fluxes 
several orders of magnitude smaller than in the previous case.

Another important piece of information could come from the annihilation
signal at the GC. Infact, apart from the clumped halo component,
the GC region may potentially be so dense that all neutralino models would 
be ruled out \cite{gon2}. This strong constrain arises in models where 
the super-massive BH at the galactic center (GC) induces a strong 
dark matter density peak called the spike. The existence of such a spike 
is strongly dependent on the formation history of the galactic center BH 
\cite{mer}. If the BH is formed adiabatically, a spike would be present 
while a history of major mergers would not allow the survival of a spike. 

In contrast to the uncertain presence of a central spike in the dark matter 
distribution, the central BH is known to induce an accretion flow of  baryonic 
matter around its event horizon. The accretion flow carries magnetic fields, 
possibly amplified to near equipartition values due to the strong compression. 
The distribution of electrons and positrons (hereafter called electrons) 
produced by neutralino annihilation at the GC would also be compressed 
toward the BH radiating through synchrotron and inverse Compton scattering 
off the photon background.

By considering the injection of electrons, combined with radiative losses and 
adiabatic compression, we find the equilibrium spatial and spectral electron 
distribution and derive the expected radiation signal.
We find that the synchrotron emission of electrons from neutralino 
annihilation range from radio and microwave energies up to the optical, in the
central more magnetized region of the accretion flow. At low frequencies, 
synchrotron self-absorption slightly reduces the amount of  radiation 
transmitted outwards. The resulting signal is stronger than the observed 
emission in the $10$ to $10^{5}$ GHz range for the case of a spiky dark matter 
profile while for a pure NFW profile the emission is below the observed values.

The paper is structured as follows in the first three paragraphs we will 
discuss the annihilation signal that could come, in the $\gamma$-ray frequency 
range, from the clumped halo of the Galaxy, while in the last three paragraphs 
we will discuss the synchrotron signal that could come, in the radio frequency
range, from the GC. We will conclude in paragraph 7.

\section{Dark Matter Clumps in the Halo}

The contribution to the diffuse $\gamma$-ray background
from $\chi\bar{\chi}$ annihilation in the clumpy halo depends on the
distribution of clumps in the halo and on the density profile of these 
clumps. Our purpose here is to investigate the wide variety of possibilities 
currently allowed by the results of simulations and suggested by some 
theoretical arguments, concerning the density profiles of dark matter 
clumps. We consider three cases: singular isothermal spheres (SIS), 
Moore et al. profiles, and NFW profiles. 

The Moore et al. and the NFW profiles are both the result of fits
to different high resolution simulations \cite{moo}.
Although there is ongoing debate over which profile 
is most accurate, it is presently believed that a realistic descriptions 
of the dark matter distribution in halos will follow a profile in the 
range defined by the Moore et al. and the NFW fits \cite{tas}. 
The dark matter density profiles can be written as follows:

\begin{equation}
\rho_{\chi,{\rm SIS}}(r) = \rho_0 \left({\frac{r}{r_0}}\right)^{-2} .
\label{SIS_profile}
\end{equation}

\begin{equation}
\rho_{\chi,{\rm Moore}}(r) = \frac{\rho_0}
{\left(\frac{r}{r_f} \right)^{3/2} \left [1+
\left(\frac{r}{r_f} \right)^{3/2} \right]}
\label{Moore_profile}
\end{equation}

\begin{equation}
\rho_{\chi,{\rm NFW}}(r) = \frac{\rho_0}
{\left(\frac{r}{r_f} \right) \left(1+
\frac{r}{r_f} \right)^2}~.
\label{NFW_profile}
\end{equation}

The SIS and Moore et al. clump density profiles are in the form given in
eqs. (\ref{SIS_profile}) and (\ref{Moore_profile}) down to a minimum radius, 
$r_{\rm min}$. 
Inside $r_{\rm min}$, neutralino annihilations are faster than the cusp 
formation rate, so that $\rho (r \le r_{\rm min}) = \rho (r_{\rm min})$ 
remains constant. To estimate $r_{\rm min}$, following \cite{ber1},
we set the annihilation timescale equal to the free-fall 
timescale, so that
\begin{equation}
r_{\rm min,SIS} = r_{0} \left[ 
\frac{\langle \sigma v \rangle_{Ann} \rho_0}
{\sqrt{G M_c} m_{\chi}} r_0^{3/2} \right]^{1/2}
\qquad
r_{\rm min,Moore} = r_{0} \left[ 
\frac{\langle \sigma v \rangle_{Ann}^2 \rho_0^2}
{G M_c m_{\chi}^2} r_f^{3} \right]^{1/3},
\label{Rmin}
\end{equation}
where $\langle \sigma v \rangle_{Ann}$ is the $\chi
\bar{\chi}$ annihilation cross-section, $G$ is the Newton constant, 
$M_c$ and $m_{\chi}$ are respectively the clump mass and the neutralino
mass. 

The fundamental parameters of the clump density profile are the density 
normalization $\rho_0$, the clump radius $r_0$ and, in the case of Moore 
et al. and NFW profiles, the clump fiducial radius $r_f$. In order to fix 
these fundamental parameters we have considered two different scenarios
(type I and II).

We have modeled the smooth galactic halo density with a NFW
density profile [Eq. (\ref{NFW_profile})], with $r_f=27$ kpc and
$\rho_0$ determined from the condition that the dark matter density at 
the Sun's position is 
$\rho_{DM}(d_{\odot}) = 6.5 \times 10^{-25} {\rm g/cm}^3$.

In the type I scenario, the radius of a clump with fixed mass is determined
by its position in the Galactic halo. More specifically, the radius of the
clump is located at the radius where the clump density equals the density 
of the Galactic (smooth) dark matter halo at the clump position (namely
$\rho_0$). The physical motivation for such a choice is to account for the
tidal stripping of the external layers of the clump while the clump is
moving in the potential of the host halo. For the NFW and Moore profiles,
the fiducial radius $r_f$ has been taken as a fixed fraction of $r_0$:
$r_f=0.1 r_0$. 

In the type II scenario the external radius of the clumps is taken to 
be their virial radius, defined in the usual way:
$r_0=r_{vir}=\left [\frac{3 M_c}{4\pi\rho_{200}} \right]^{1/3}$,  
where $\rho_{200}$ is $200$ times the critical density of the Universe
$\rho_c=1.88\times 10^{-29}h^2$ g/$cm^3$ (we have assumed $h=0.7$ everywhere).
In this scenario, following \cite{ull}, we have introduced 
the concentration parameter defined as $\xi=\frac{r_0}{r_{-2}}$,
where $r_{-2}$ is the radius at which the effective logarithmic slope of 
the profile is $-2$, set by the equation 
$ \frac{1}{\rho_0}\frac{d}{d r}r^2\rho(r) = 0$. 
The mass dependence of the concentration parameter used in our calculations
is taken from \cite{wec}.

The definite trend
is that smaller clumps have larger concentration parameter, reflecting the
fact that they are formed at earlier epochs, when the Universe was denser. 
In general, the concentration parameter has also a dependence on the redshift 
at which the parameter is measured. We are not interested here in such 
dependence, since we only consider what happens at the present time (zero 
redshift). The normalization constant in the clump density profile 
$\rho_0$ is fixed by the total clump mass $M_c$.

In the case of the NFW density profile $r_f=r_{-2}$, while in the case of the
Moore et al. density profile $r_f=r_{-2}/0.63$ \cite{ull}. In 
terms of concentration parameters $\xi_{NFW}=0.63 \xi_{Moore}~$. 
Using the concentration parameter as in \cite{wec} for NFW
clumps, we can estimate the stripping distance, which we define as the typical 
distance from the galactic center where the density of a clump of fixed 
mass and the density in the smooth DM profile are equal at the fiducial 
radius $r_f$ of the clump. In other words, at the stripping distance the 
layers of the clump outside the fiducial radius will have been stripped off.
From this estimate it is easy to see that most clumps
in the inner parts of a host galaxy are stripped off of most of their material,
so that this region has no clumpy structure. Another way of seeing this
phenomenon is that the clumps that are able to reach the central part of
the Galaxy are effectively merged to give the observed smooth dark matter
profile. 

The exception to this conclusion may be represented by low mass
clumps, which are more concentrated (denser cores) and may then penetrate
deeper. Numerical simulations show the disappearance of large clumps in 
the centers of galaxy size halos, but they cannot resolve the smaller denser
clumps that may eventually make their way into the core of the galaxy.
For simplicity, in our calculations we assume that the inner 10 kpc of 
the Galaxy have no clumps at all.

\begin{figure}[t!]
\begin{tabular}{ll}
\epsfxsize=6cm   
\epsfbox{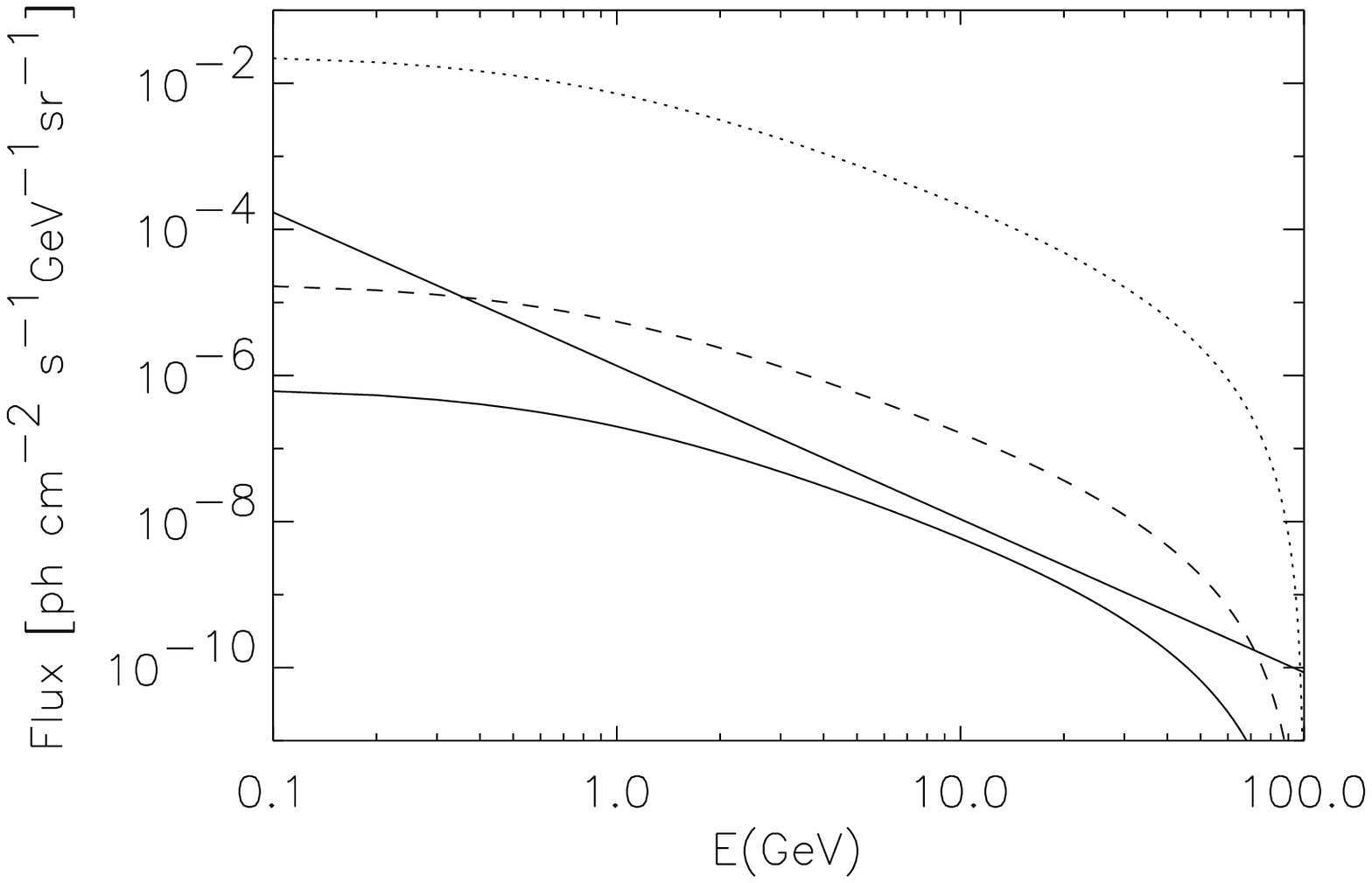} &
\epsfxsize=6cm   
\epsfbox{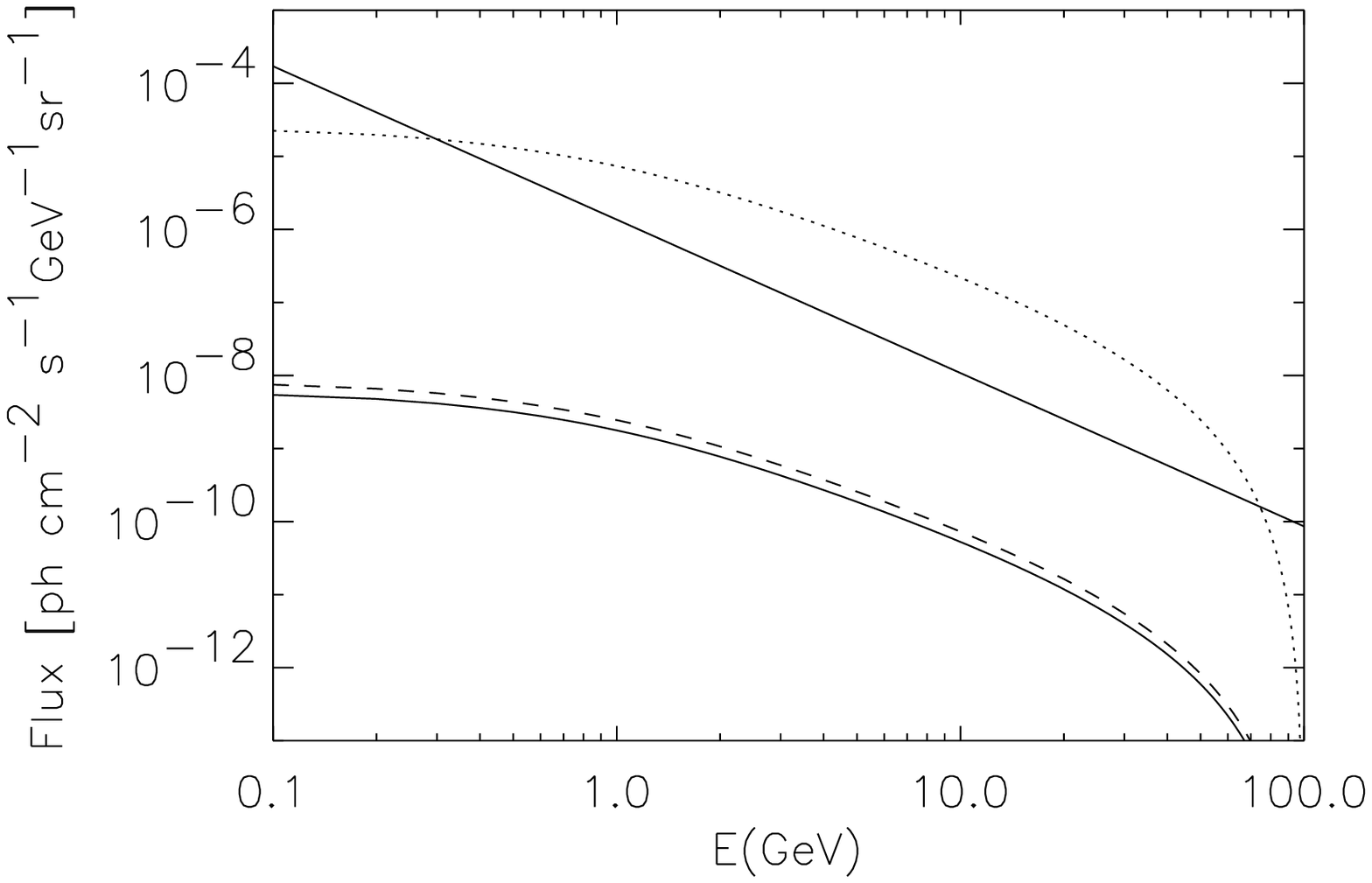} \\
\end{tabular}
\caption{Flux of $\gamma$-rays in units of 
(GeV cm$^2$ s sr)$^{-1}$ arriving on Earth averaged in all directions for 
$m_{\chi}=$ 100 GeV, $\langle \sigma v \rangle_{Ann}=3\times 10^{-27}$ cm$^3$/s
and $M_{c,min}=10^5$ $M_{\odot}$. SIS density profile (dotted line), Moore
et al. density profile (dashed line) and NFW density profile (continuous line). 
Also shown is the EGRET data on extragalactic diffuse $\gamma$-ray background
(left panel: first scenario; right panel: second scenario). 
\label{fig:gamflux}
}
\end{figure}

\section{Gamma-ray emission in Neutralino Annihilation}

In order to determine the $\gamma$-ray emission from the DM clumps, as already
pointed out in the previous section, the distribution of clumps in the Galaxy 
is needed. The probability distribution function of clumps
with a given mass and at a given position can be fitted from numerical 
simulations. In the present paper we follow \cite{blaSheth} adopting a 
spatial (as a function of the distance $d$ from the galactic center) and 
mass distribution of the clumps reflecting the following expression: 
\begin{equation}
N_c (d,M_c) = N_{c,0} \left(\frac{M_c}{M_H}\right)^{-\alpha}
\left [1+\left (\frac{d}{d_{sc}}\right)^2 \right]^{-3/2},
\label{clumpdist}
\end{equation}
where $N_{c,0}$ is a normalization constant and $d_{sc}$
is the scale radius of the clump distribution\footnote{In this paper we have 
assumed $d_{sc}=10$ Kpc as in \cite{blaSheth}.}. 
Simulations find $\alpha\simeq 1.9$  and a halo like that of our Galaxy, with 
$M_H\simeq 2\times 10^{12}~M_\odot$, contains about $500$ clumps with 
mass larger than $10^8~M_{\odot}$ \cite{ghi}.
The $\gamma$-ray flux per unit solid angle and per unit energy along a fixed 
line of sight in the $(\theta,\phi)$ direction can be computed as 
\begin{equation}
\Phi_{\gamma}(E_{\gamma},\theta,\phi)=\frac{1}{4\pi}\int_0^{s_{max}}
ds \int_{M_{min}}^{\zeta M_H} N_c(d(s),M) N_{\gamma} ~dM
\label{gammaflux}
\end{equation}
where $d(s)=\sqrt{s^2-2sd_{\odot}cos\theta+d_{\odot}^2}$ is the 
distance of a generic point on the line of sight from the galactic center
(with $\theta$ the angle between the direction $s$ and the axis Sun-galactic
center), $M$ is the clump mass, $\zeta M_H$ is the maximum allowed mass 
for DM clumps in the Halo (we have used $\zeta=0.01$) and $N_{\gamma}$ 
is the total number of photons emitted per unit time and energy by a DM 
clump of mass $M$. This quantity, depending on the scenario chosen for the
clump density profile, may depend or not on the distance $d(s)$ of the 
considered point from the galactic center. In the first scenario, where the 
normalization of the clump density is related to the smooth Halo density, 
one has $N_{\gamma}=N_{\gamma}(M,d,E_{\gamma})$, 
while in the second scenario 
$N_{\gamma}=N_{\gamma}(M,E_{\gamma})$. 
One should remember that Eq. (\ref{gammaflux}) is an average over all
possible realizations of a halo with its clumpy structure. Fluctuations
around this value may be present due to the accidental proximity of few
clumps in the specific realization that we happen to experience in our
Galaxy.

We assume that neutralinos mainly annihilate into quark-antiquark pairs,
which seems confirmed by more detailed calculations carried out in specific
supersymmetric scenarios \cite{ull}. Actually, it is easy to see that
even when neutralinos annihilate into pairs of $W^{+}W^{-}$ or $Z^0 Z^0$,
the end result of the decay chain is dominated by quarks and antiquarks,
hadronizing mainly into pions (roughly 1/3 neutral pions and 2/3 charged 
pions) and their spectral shape is identical to that of direct 
quark-antiquark production. In fact, each $W$ or $Z$ boson would have a Lorentz
factor $\sim m_\chi/M_{Z,W}$ (if the neutralino mass is large enough to 
allow the production of a $W$ or $Z$ pair). In the rest frame of the boson,
the maximum energy of the particles generated in the decay is $M_{Z,W}$, so
that in the laboratory frame the spectrum has a cutoff at $m_\chi$, as in
the case of direct quark production in the neutralino annihilation. The
spectrum is also left unchanged.

The number of photons produced with energy $E_{\gamma}$ in a single 
$\chi \bar{\chi}$ annihilation can be written as follows:
\begin{equation}
\frac{dN_{\gamma}}{d E_{\gamma}}=\int_{E_{\pi,min}}^{E_{\pi,max}}
dE_{\pi} P(E_{\pi},E_{\gamma})\frac{dN_{\pi}}{dE_{\pi}}
\end{equation}
where $P(E_{\pi},E_{\gamma}) = 2 (E_{\pi}^2 - m_{\pi}^2)^{-1/2}$ is 
the probability per unit energy to produce a $\gamma$-ray with energy
$E_{\gamma}$ out of a pion with energy $E_{\pi}$. For the pion 
fragmentation function we assume the functional form introduced by Hill
\cite{hill}:
\begin{equation}
\frac{dN_{\pi}}{dE_{\pi}} = \frac{1}{m_{\chi}}\frac{15}{16} 
x^{-3/2}(1 - x)^2 \, 
\label{piFF}
\end{equation}
with $x = E_{\pi}/m_{\chi}$, $E_{\pi,\rm max} = m_{\chi}$ and 
$E_{\pi,\rm min} = E_{\gamma}+m_{\pi}^2/4 E_{\gamma}$. Finally,
\begin{equation}
\frac{dN_{\gamma}}{dE_{\gamma}} = \frac{5}{4 m_{\chi}} \int_{x_{\rm
m}}^{1} dx \frac{(1 - x)^2}{x^{3/2} (x^2 - \eta^2)^{1/2}}~,
\label{gamspec}
\end{equation}
where  $\eta = m_{\pi}/m_{\chi}$, and $x_{\rm m} = E_{\gamma}/m_{\chi} +
m_{\chi} \eta^2/4 E_{\gamma}$. 

The neutralino annihilation rate per unit volume is, 
$\Gamma_{\chi\bar{\chi}}(r,M,E_{\gamma}) = \rho_{\chi}^2(r) 
\langle \sigma v \rangle_{Ann} /m_{\chi}^2$,
therefore the $\gamma$-ray emissivity $j_{\gamma}(r,M,E_{\gamma})$ 
associated to the single clump of mass $M$ is obtained by multiplying Eq.
(\ref{gamspec}) by $\Gamma_{\chi\bar{\chi}}$. The number of $\gamma$-rays 
produced per unit time and per unit energy in a single DM clump of mass 
$M$ is then
\begin{equation}
N_{\gamma}(M,E_\gamma)=
\int_0^{r_0} dr 4\pi r^2 j_{\gamma}(r,M,E_\gamma)~.
\label{gammarate}
\end{equation}

\section{Gamma Ray Emission from Clumps}

In this section we present the results of our calculations of the 
$\gamma$-ray emission from dark matter annihilation in the halo of
our Galaxy, including both the smooth and clumped components introduced
above. We detail the description of these results for the two scenarios 
(type I and II) of spatial distribution of the clumped component.
We expect the type II scenario to result in a quite weaker signal than
that obtained in the type I scenario, because the recipe for the type
I clumps implies much stronger concentration. In the type II scenario the 
$\gamma$-ray signal from clumps overcomes the $\gamma$-ray flux from the smooth
dark matter distribution only in the direction of the galactic anticenter
and only for SIS and Moore profiles for the dark matter distribution
inside the clumps. 

On the other hand, in the type II scenario the contribution of the clumped 
component to the diffuse $\gamma$-ray flux from the galactic halo is several 
orders of magnitude above the smooth halo component, for any of the three 
density profiles considered above. This impressive difference in the 
predictions is symptomatic of a large uncertainty in the physics involved
in the formation and survival of dark matter substructures.

In Fig. \ref{fig:gamflux} we plot the flux of $\gamma$-rays 
in units of (GeV cm$^2$ s sr)$^{-1}$ arriving on Earth averaged in all 
directions for $m_{\chi}=$100 GeV and with 
$\langle \sigma v \rangle_{Ann} = 3 \times 10^{-27}$ cm$^3$/s. The curves refer
to the $\gamma$-ray flux due to the full dark matter profile, made
of the smooth and clumped components. For the type II scenario, the 
$\gamma$-ray flux contributed by the clumped component is comparable to 
the contribution of the smooth dark matter profile, while for the 
type I scenario, the clumpy component is overwhelmingly larger than that 
due to the smooth component. 

The fluxes plotted in Fig. \ref{fig:gamflux} are 
obtained choosing the minimum clump mass of $M_{c,min}=10^5 M_{\odot}$,
but the dependence of these fluxes on the value of $M_{c,min}$ is 
only logarithmic.
In both figures the dotted, dashed and solid lines correspond to SIS,
Moore and NFW clump density profiles respectively.

Also shown are the EGRET data (straight line) on the extragalactic diffuse 
$\gamma$-ray background  which can be fitted from 30 MeV to $\sim$ 30 GeV by 
\cite{sre} ${\frac{dN_{eg}}{d\Omega dE}}=
1.36\times 10^{-6} \left(\frac{E}{{\rm GeV}}\right)
^{-2.10} {\rm GeV}^{-1}{\rm cm}^{-2}{\rm s}^{-1}{\rm sr}^{-1}$.
Depending on the density profile the fluxes have different scalings with 
the neutralino parameters $m_{\chi}$ and $\langle \sigma v \rangle_{Ann}$: 
\begin{equation}
\Phi_{SIS}\propto \langle \sigma v \rangle_{Ann}^{1/2} m_{\chi}^{-5/2}
\qquad
\Phi_{NFW,Moore}\propto \langle \sigma v \rangle_{Ann} m_{\chi}^{-3}~,
\label{scalings}
\end{equation}
these scalings are the same for the type I and II scenarios.

It is clear from Fig. \ref{fig:gamflux} that the 
comparison between our predictions and the observed diffuse background 
is meaningful only for the type I scenario, with highly concentrated
clumps. In the second scenario, the fluxes are too low, with the 
exception of the case in which the density profile is the SIS one.
For the other cases the region of parameters that can be constrained 
is already ruled out from accelerator experiments $m_{\chi}\ge 50$ GeV 
\cite{Hagiw} and from theoretical arguments 
$\langle \sigma v \rangle_{Ann} \le 10^{-26}$cm$^{3}$/s \cite{Baltz}.

\begin{figure}[t!]
\epsfxsize=8cm   
\epsfbox{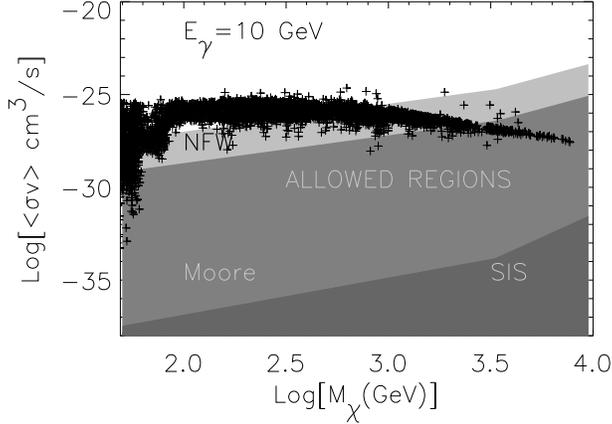}
\caption{Limits on the neutralino parameter space in the first scenario
for clump density profile. 
\label{fig:bound}} 
\end{figure}

The situation is different for the type I scenario. As anticipated above,
all the density profiles imply $\gamma$-ray fluxes comparable to or largely
in excess of the EGRET observations of the diffuse $\gamma$-ray background.
These observations may therefore be used as a tool to extract severe
constraints on the neutralino parameter space, which unfortunately are 
restricted to the type I scenario. Using the scalings given above 
(cfr. Eq. (\ref{scalings})), we can determine the regions of the parameter 
space which are ruled out by our calculations.

Using EGRET data at 10 GeV, we find that SIS clumps in our halo are ruled 
out in the region 
$\langle \sigma v \rangle_{Ann} \ge 2 \times 10^{-35} 
(m_{\chi}/{\rm 100 GeV})^{2} {\rm cm}^3/{\rm s}$ 
for $50 GeV \leq m_\chi \leq 3 TeV$, and
$\langle \sigma v \rangle_{Ann} \ge 6 \times 10^{-40} 
(m_{\chi}/{\rm 100 GeV})^{5} {\rm cm}^3/{\rm s}$ 
cm$^3$/s for $m_{\chi} \ge 3$ TeV. 

Moore et al. clumps are also strongly constrained: the 10 GeV EGRET data 
require that 
$\langle \sigma v \rangle_{Ann} \le 2.5 \times 10^{-28} 
(m_{\chi}/{\rm 100 GeV})^{3/2} {\rm cm}^3/{\rm s}$ 
for $50 GeV \leq m_\chi \leq 3 TeV$, and 
$\langle \sigma v \rangle_{Ann} \le  
10^{-30} (m_{\chi}/{\rm 100 GeV})^{3} {\rm cm}^3/{\rm s}$
for $m_{\chi} \ge 3$ TeV.

If we extrapolate the EGRET measurement of the extragalactic diffuse 
$\gamma$-ray background to 100 GeV, the bounds get tighter: for $m_{\chi}$ 
between 50 GeV and 30 TeV, the flux from clumps is below the EGRET data if 
$\langle \sigma v \rangle_{Ann} \le 6 \times 10^{-29} (m_{\chi}/{\rm 100
GeV})^{3/2}$  cm$^3$/s, while for $m_{\chi} \ge 30$ TeV, the region 
$\langle \sigma v\rangle_{Ann} \le  10^{-32} (m_{\chi}/{\rm 100 GeV})^{3}$ 
cm$^3$/s is allowed.

The NFW clumps in the type I scenario are the ones that are more weakly 
constrained. The bounds that can be placed by EGRET at 10 GeV are as follows:
if $m_{\chi}$ is between 50 GeV and 3 TeV, the allowed region is defined by
$\langle \sigma v \rangle_{Ann} \le  10^{-26} (m_{\chi}/{\rm 100
GeV})^{3/2}$  cm$^3$/s. For $m_{\chi} \ge 3$ TeV, the allowed region is instead
$\langle \sigma v \rangle_{Ann} \le 6 
\times 10^{-29} (m_{\chi}/{\rm 100 GeV})^{3}$ 
cm$^3$/s. If we extrapolate EGRET data up to 100 GeV, the bounds become 
as follows: for $m_{\chi}$ between 50 GeV and 30 TeV, one must have
$\langle \sigma v \rangle_{Ann} \le 3 \times 10^{-27} (m_{\chi}/{\rm 100
GeV})^{3/2}$  cm$^3$/s while for $m_{\chi} \ge 30$ TeV, the allowed region
becomes 
$\langle \sigma v\rangle_{Ann} \le 5 
\times 10^{-31} (m_{\chi}/{\rm 100 GeV})^{3}$  
cm$^3$/s. All these bounds are shown in Fig. \ref{fig:bound}. The scatter plot
reported in Fig. \ref{fig:bound} represents all the admitted values of 
$(m_\chi,\langle \sigma v \rangle_{Ann})$ allowed by SUSY 
models \cite{Baltz}. 

\section{Dark Matter and Magnetic Field at the GC}

The possibility of detecting neutralino annihilation trough the $\gamma$-ray 
emission, as discussed in the previous sections, is not the only one. 
Another appealing possibility is related to the synchrotron emission of 
electron-positron pairs produced by neutralino annihilation at the GC. 
In this section we will discuss the major characteristics of the DM density
profile at the GC, taking also into account the effects on the GC magnetic
field of the advection flow produced by the central BH.

The spatial distribution of dark matter in galactic halos is still a matter 
of much debate (see, e.g., \cite{tas}). Numerical simulations suggest that 
collisionless dark matter forms cuspy halos while some observations argue for 
a flat inner density profile \cite{sal}. The standard numerical dark matter 
halo is the NFW profile \cite{nfw} which is expected to be universal. 
However, recent simulations have found more cuspy halos \cite{moo} as well 
as shallower profiles \cite{pow}. At present, it is not clear if there is a 
universal dark matter halo profile, but the NFW profile seems to represent 
well the range of possibilities. Therefore, we assume that the NFW profile 
describes well the dark matter in our Galaxy and we will use the NFW profile
with the parameters introduced in \S 2.

There is now growing evidence for the presence of a supermassive BH at 
the GC, with mass $\sim 2\times 10^6 M_\odot$. 
In fact most galaxies seem to have central black holes with comparable or even 
larger masses. The presence of a BH can steepen the density profile of dark 
matter by transforming the cusp at the GC
into a spike of dark matter \cite{gon1}. The density profile of the spike 
region, where the gravitational potential is dominated by the BH is described 
by 
\begin{equation}
\rho'_{sp}(r)=\alpha_\delta^{\delta_{sp}-\delta}
\left( \frac{M}{\rho_\odot R_\odot^3}\right)^{(3-\delta)(\delta_{sp}-\delta)}
\rho_\odot g(r) \left(\frac{R_\odot}{r}\right)^{\gamma_{sp}}.
\label{eq:spike}
\end{equation}
Here, $\delta$ is the slope of the density profile of dark matter in the inner 
region ($\delta=1$ for a NFW profile), and 
$\delta_{sp}=(9-2\delta)/ (4-\delta)$.
The coefficients $\alpha_\delta$ and $g(r)$ can be calculated numerically as 
explained in detail in \cite{gon1}. It is possible to identify a spike 
radius $R_{sp}$ where the spike density profile given in Eq. (\ref{eq:spike}) 
matches the NFW dark matter profile. In other words, at $R_{sp}$ the 
gravitational potential is no longer dominated by the central BH.

Neutralino annihilations affect the density profile in the spike by generating 
a flattening where the annihilation time becomes smaller than the age of the 
BH. This effect produces a constant neutralino density given by
$\rho_{core} = \frac{m_\chi}{\langle \sigma v \rangle_{Ann} 
t_{BH}}$, where $t_{BH}$
is the BH age. The effect of annihilations on the spike density profile 
can be written as
\begin{equation}
\rho_{sp}(r) = \frac{\rho'_{sp}(r) \rho_{core}}{\rho'_{sp}(r) + \rho_{core}},
\end{equation}
which accounts for the flattening in the central region.

Several dynamical effects may weaken or destroy the spike in the GC 
\cite{jun,mer}, depending on the history of formation of the central BH.
If the spike is not formed or gets destroy, the central region should be 
described by the cuspy profile such as in the NFW case. Here we consider both 
cases and show that the observed emission is stronger than the predictions for 
a NFW cusp, while the spike generates signals well above the observations.

In order to determine the synchrotron emission produced by the electrons that 
come from neutralino annihilations it is necessary to determine the magnetic
field present in the GC region. In what follows we will determine the 
magnetic field strength assuming the equipartition between magnetic field 
energy and the kinetic pressure due to the accretion flow inside the 
central BH.

We model the accretion flow of gas onto the BH at the center of our 
Galaxy following a simple approach described in \cite{melia}. More detailed 
models of the accretion flow around the BH lead to corrections which are 
negligible when compared to the uncertainties in the dark matter distribution.
In the model we adopt, the BH accretes its fuel from a nearby molecular cloud,
located at about $0.01$ pc from the BH. The accretion is assumed to be
spherically symmetric Bondi accretion with a rate of mass accretion of $\dot M=
10^{22} \dot M_{22}~\rm g~s^{-1}$. The accretion onto the BH occurs with a 
velocity around the free-fall velocity, such that 
\begin{equation}
v(r)=\sqrt{2 G M_{BH}/r} = c \left(\frac{R_g}{r}\right)^{1/2}
\end{equation}
where $R_g=2GM_{BH}/c^2=7.4\times 10^{11} 
(M_{BH}/2.5\times 10^6 M_\odot)~\rm cm$ is the gravitational radius of the BH 
and $M_{BH}$ is the BH mass. Therefore,
\begin{equation}
v(r)= 1.46 \times 10^8
\left(\frac{M_{BH}}{2.5\times 10^6 M_\odot}\right)^{1/2} 
\left(\frac{r}{0.01 \rm pc}\right)^{-1/2}~\rm cm~s^{-1}.
\end{equation}
Mass conservation then gives the following density profile:
\begin{equation}
\rho(r) = \frac{\dot M}{4 \pi r^2 v(r)} =  \frac{\dot M}{4 \pi R_g^2 c}
\left(\frac{r}{R_g}\right)^{-3/2} \ ,
\end{equation}  
such that 
\begin{equation}
\rho(r) =  5.6\times 10^{-21} \dot M_{22}
\left(\frac{M_{BH}}{2.5\times 10^6 M_\odot}\right)^{-1/2}
\left(\frac{r}{0.01 \rm pc}\right)^{-3/2}~\rm g~cm^{-3} \ .
\end{equation}  

Following \cite{melia}, we assume that the magnetic field in the
accretion flow achieves its equipartition value with the kinetic pressure, 
namely $\rho v^2/2 = B(r)^2/8\pi$. With this assumption,
\begin{equation}
B_{eq}(r)=\frac{\sqrt{\dot M c}}{R_g}
\left(\frac{r}{R_g}\right)^{-5/4}
=3.9\times 10^4 \dot M_{22} M_{BH}^{1/4} \left(\frac{r}{0.01\rm pc}
\right)^{-5/4} \quad \mu {\rm G}.
\label{eq:magnetic}
\end{equation}

It is believed that magnetic fields in the accretion flow will in general 
reach the equipartition values described in Eq. (\ref{eq:magnetic}). 
However, smaller fields may be reached if the equipartition is prevented 
somehow. In what follows we will always assume the equipartition field 
in deriving the synchrotron signal from electrons produced by neutralino 
annihilation. 

\section{Neutralino Annihilation at the GC}

Following our discussion of section \S 3 we will assume that neutrino
annihilation channel is dominated by quark-antiquark production, with 
a large production of pions described, as already discussed in \S 3, by the
Hill \cite{hill} spectrum.

The spectrum of electrons (and positrons) from the $\pi^\pm$ decays is 
calculated by convoluting the spectrum of pions and muons. For relativistic 
electrons the electron spectrum reads
\begin{equation}
W_e(E_e)=\int_{{\rm max}(E_e,m_\mu)}^{m_\chi} dE_\mu
\int_{E_\pi^{{\rm min}}}^{E_\pi^{{\rm max}}}
d E_\pi W_\pi(E_\pi) \frac{m_\pi^2}{m_\pi^2-m_\mu^2}
\frac{1}{\sqrt{E_\pi^2-m_\pi^2}}\frac{dn_e(E_e,E_\mu,E_\pi)}{dE_e},
\end{equation} 
where, neglecting the muon polarization, we get 
\begin{equation}
\frac{dn_e(E_e,E_\mu,E_\pi)}{dE_e}=\frac{1}{E_\mu\beta}
\left\{ \begin{array}{ll}
2\left [\frac{5}{6}-\frac{3}{2}\epsilon^2+\frac{2}{3}\epsilon^3\right ]
&{\rm if} ~~~\frac{1-\beta}{1+\beta}\le \epsilon \le 1 \\
\frac{4\epsilon^2\beta}{(1-\beta)^2}
\left [3-\frac{2}{3}\epsilon\left(\frac{3+\beta^2}{1-\beta}\right)^2\right ]
&{\rm if} ~~~0\le\epsilon\le\frac{1-\beta}{1+\beta},
\end{array}
\right.
\end{equation}
with $\epsilon=\frac{2}{1+\beta}\frac{E_e}{E_\mu}$. Here $\beta$ is the pion
speed and $E_e$ and $E_\mu$ are the total energies of electrons and muons
respectively. The two limits of integration 
$E_{\pi}^{{\rm min}}(E_\mu)$ and 
$E_{\pi}^{{\rm max}}(E_\mu)$ can be derived by inverting the following 
equations:
\begin{equation}
E_\mu\le \frac{E_\pi}{2m_\pi^2}
\left [m_\pi^2 (1+\beta)+m_\mu^2 (1-\beta)\right ]
\end{equation}
\begin{equation}
E_\mu\ge \frac{E_\pi}{2m_\pi^2}
\left [m_\pi^2 (1-\beta)+m_\mu^2 (1+\beta)\right ]~.
\end{equation}

Finally, the injection of new electrons, produced in neutralino annihilation,
at the distance $r$ from the BH and at energy $E$ can be written as 
$Q(E,r)=(1/2)(\rho_{DM}(r)/m_\chi)^2 W_e(E) \langle \sigma v\rangle_{Ann}$,
where the density of dark matter has the profile $\rho_{DM}(r)$ discussed 
in \S 5 (NFW or spike).

The spectrum of particles at a position $r$ in the accretion flow around 
the BH is the result of the injection on newly produced electrons at the 
same position, radiative losses of these electrons and the adiabatic 
compression that may enhance their momentum while they move inward.
Here we neglect spatial diffusion, which occurs on larger time scales.

The transport equation including all these effects can be written as follows:
\begin{equation}
v(r)\frac{\partial f}{\partial r} - \frac{1}{3 r^2}\frac{\partial}{\partial r}
\left[r^2 v(r)\right]~p~ \frac{\partial f}{\partial p}+
\frac{1}{p^2}\frac{\partial}{\partial p}\left[ p^2 \dot p(r,p) f\right]
=Q(r,p),
\label{eq:transport}
\end{equation}
where $f(r,p)$ is the equilibrium distribution function of electrons injected
according with $Q(r,p)$, and losing energy radiatively as described by the
function $\dot p(r,p)= dp(r,p)/dt$. Here $v(r)=-c(r/R_g)^{-1/2}$ is the inflow 
velocity. The equation can be solved analytically if the electrons remain 
relativistic everywhere in the fluid. The assumption of relativistic electrons
can be safely used as discussed in \cite{noi2}. In Eq. (\ref{eq:transport}), 
the term
\begin{equation}
\dot p_{ad} = -\frac{1}{3} p \nabla v(r) = 
-\frac{1}{3 r^2} p \frac{\partial}{\partial r}\left[r^2~v(r)\right]
\end{equation}
describes the rate of change of momentum of a particle at the position $r$ 
due to adiabatic compression in the accretion flow. The rate of adiabatic 
momentum enhancement should  be compared with the rate of losses due to 
synchrotron emission:
\begin{equation}
\dot p_{syn}(r,p)=\frac{4}{3}\sigma_T \frac{B^2(r)}{8\pi} \gamma^2,
\label{eq:syn}
\end{equation}
where $\sigma_T$ is the Thomson cross section and $\gamma$ is the Lorentz
factor of the electron. The magnetic field $B_{eq}(r)$ depends on $r$ as 
described in Eq. (\ref{eq:magnetic}).

In order to solve the transport equation, Eq. (\ref{eq:transport}), we discuss 
different loss processes using the equipartition field $B_{eq}$. We first 
consider synchrotron losses ($\dot p_{syn}$), followed by inverse Compton 
losses ($\dot p_{ICS}$) and synchrotron self-compton scattering 
($\dot p_{SSC}$).
The rate of synchrotron losses from Eq. (\ref{eq:syn}) is given by:
$\dot p_{syn}(r,p) ~ c = 1.6 \times 10^{-18} \left(\frac{r}{0.01\rm pc}
\right)^{-5/2} ~ \gamma^2 ~ \rm erg ~ s^{-1}$.
In the Thomson regime, losses due to Inverse Compton Scattering (ICS) off a 
photon background with energy density $U_{ph}$ has the following form
$\dot p_{ICS}(r,p)=\frac{4}{3}\sigma_T U_{ph} \gamma^2$,
ICS  dominates synchrotron losses only if  $B^2 >  8\pi U_{ph}$. 
Assuming that $U_{ph}$ is independent of $r$ (i.e. a fixed photon background) 
ICS becomes important at large radii ($\sim 0.01 \rm pc$) and only if 
$U_{ph}\ge 10^4 \rm eV ~ cm^{-3}$. Such a strong photon background is 
unlikely to be present at the GC region. For comparison, the CMB radiation 
has $U_{CMB}\approx 0.25 \rm eV ~ cm^{-3}$ while the optical background has 
$U_{opt}\approx 1 \rm eV ~ cm^{-3}$. If ICS off a fixed background is not 
dominant at  large radii, it becomes even less important as small radii when 
compared to synchrotron losses. Consequently, we safely neglect the role 
of ICS off photons of fixed photon backgrounds.

The electrons, radiating in the strong magnetic field near the BH generate a 
photon background that can become quite intense. The rate of losses due to ICS 
of electrons off the photons generated through synchrotron emission by 
the same electrons is
$\dot p_{SSC}(r,p)=\frac{4}{3}\sigma_T U_{ph}^{syn}(r) \gamma^2$,
where the photon energy density generally depends on the radius $r$. The 
photon density $U_{ph}^{syn}(r)$ is a nonlinear function of the distribution 
$f(r,p)$. In other words, the term $\dot p$ in Eq. (\ref{eq:transport}) 
depends in turn on $f(r,p)$ when synchrotron self-Compton scattering is 
included. If this contribution cannot be neglected, an analytical solution 
of the transport equation becomes unattainable.

\begin{figure}[t!]
\begin{tabular}{ll}
\epsfxsize=6.0cm   
\epsfbox{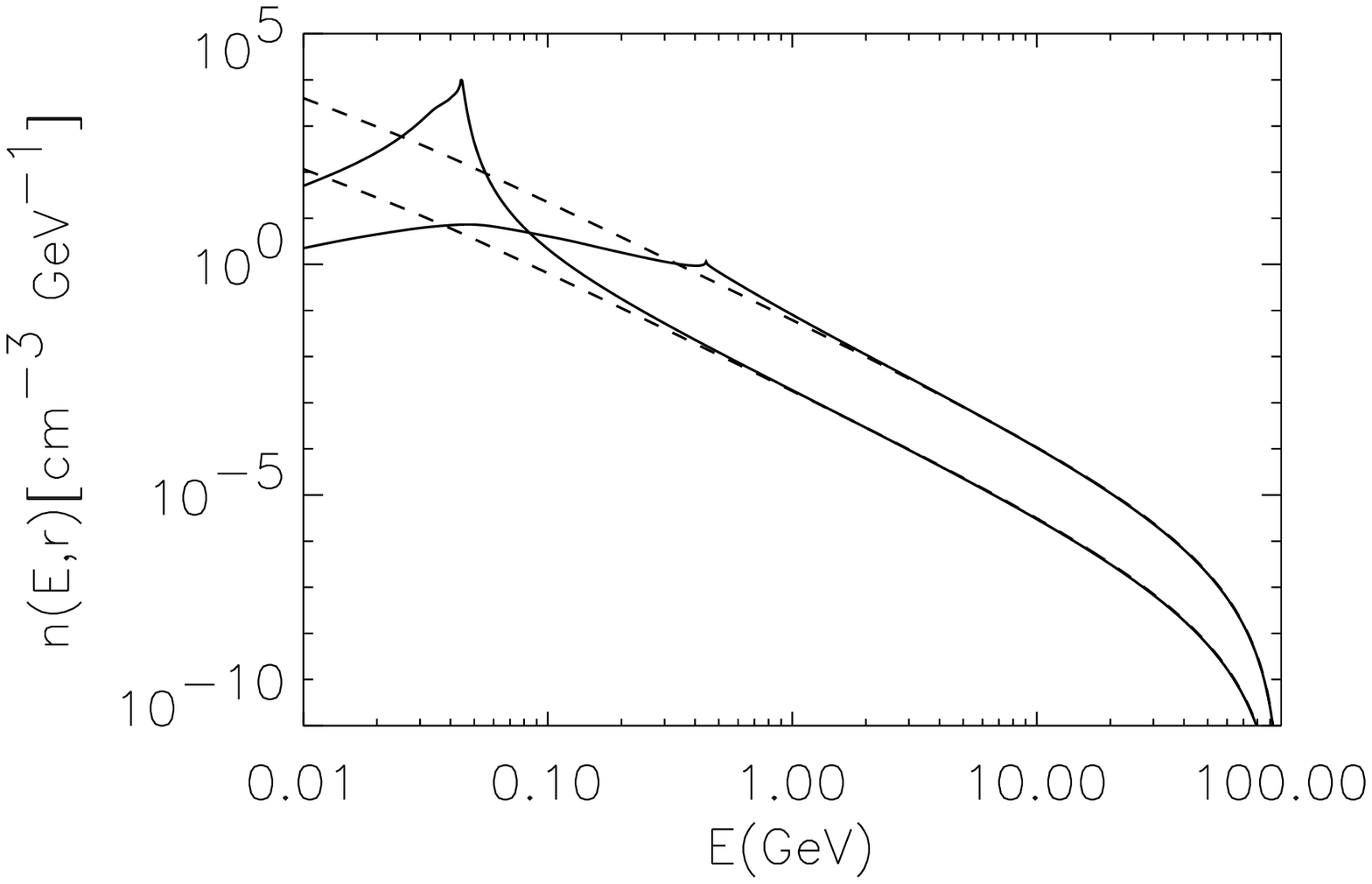} &
\epsfxsize=6.0cm   
\epsfbox{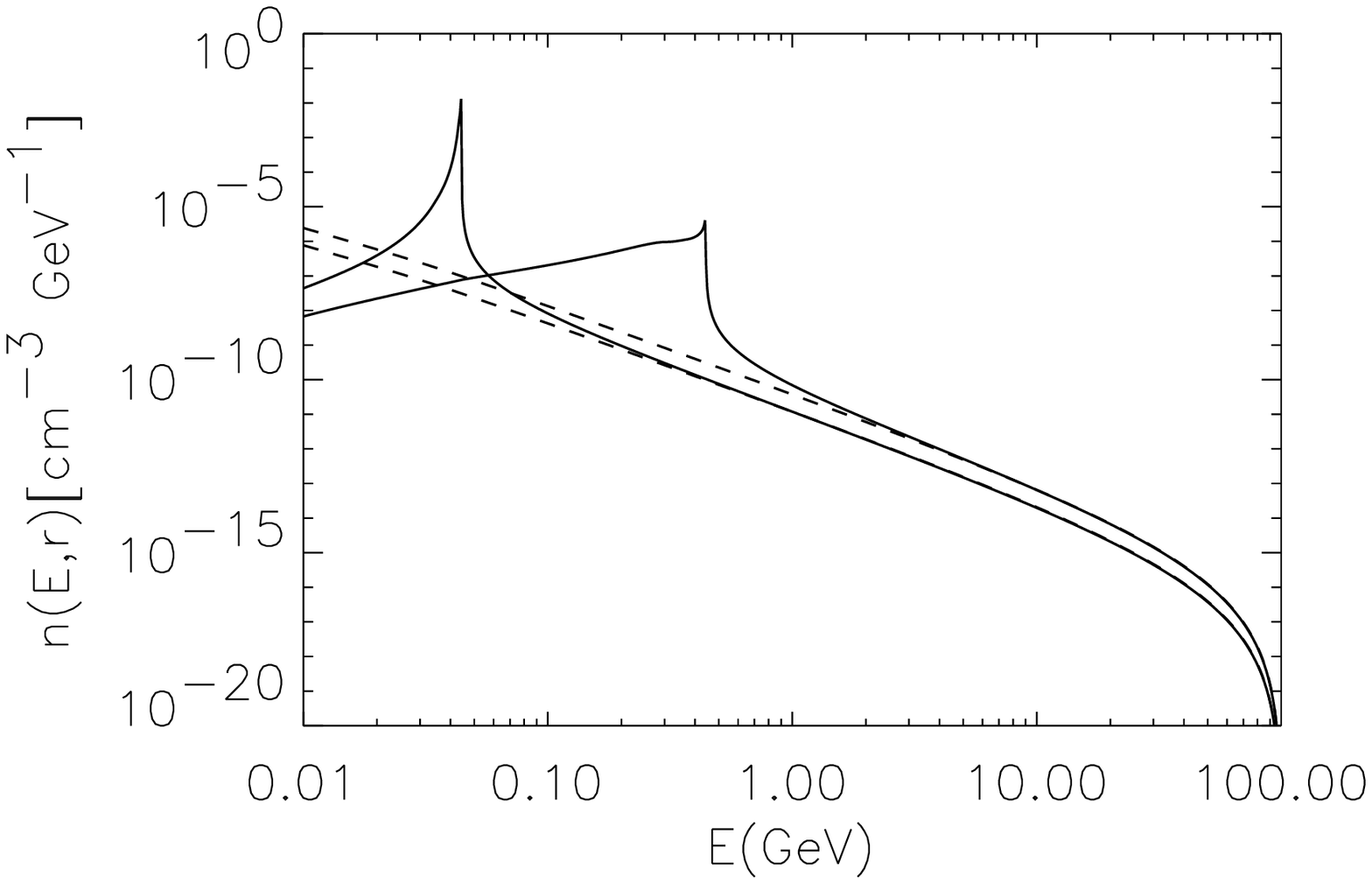} \\
\end{tabular}
\caption{Electron density per unit energy as a function of energy
(solid lines) at $r=10^{3}R_g$ (upper curve) and $r=10^{4}R_g$ 
(lower curve). Superimposed (dashed curves), we plot the function 
$n(E,r)$ obtained without the effect of advection. The left panel
illustrates the spike case, while the right panel applies to the 
NFW case.}  
\label{fig:nEr}
\end{figure}

Given the distribution function $f(r,p)$, one can calculate the synchrotron 
emissivity  $j(\nu,r)$ (energy per unit volume, per unit frequency, per unit 
time). The photon energy density at the position $r$ is then proportional 
to the integration over all lines of sight of the emissivity, with the 
possible synchrotron self-absorption taken into account at each frequency. 
However, the distribution function $f(r,p)$ is not known {\it a priori}, and 
the problem becomes intrinsically nonlinear. The approach that we follow here 
is start with neglecting  synchrotron self-Compton scattering and check 
{\it a posteriori} whether the assumption is correct in the situation at hand.

An analytic solution of the transport equation can be derived when 
$\dot p(r,p)$ is dominated by synchrotron losses as in Eq. (\ref{eq:syn}).  
In this case, the equation admits the following analytical solution:
\begin{equation}
f(r,p) = \frac{1}{c}\left(\frac{r}{R_g}\right)^{-2}
\int_r^{R_{acc}} d R_{inj}
\left(\frac{p_{inj}}{p}\right)^{4}
\left(\frac{R_{inj}}{R_g}\right)^{5/2} Q(R_{inj},p_{inj})~.
\label{eq:trans_sol}
\end{equation}

The function $p_{inj}=p_{inj}[p,r,R_{inj}]$ corresponds to the injection 
momentum of an electron injected at the position $R_{inj}$ that arrives at the 
position $r$ with momentum $p$. This injection momentum can be obtained by 
inverting, with respect to $p_{inj}$ the solution of the equation of motion of 
the electron, in the presence of adiabatic compression and radiative losses:
\begin{equation}
\frac{dp}{dr} = \frac{k_0}{c}\left(\frac{r}{R_g}\right)^{-2} p^2
-\frac{1}{2 R_g}~p~\left(\frac{r}{R_g}\right)^{-1}.
\end{equation}
The solution of this equation, with initial condition
$p[r=R_{inj},p_{inj},R_{inj}]=p_{inj}$ is
\begin{equation}
p[r,p_{inj},R_{inj}] = p_{inj} \left[
\frac{2 k_0}{3 c} \frac{R_g^2}{r}p_{inj} \left[1-\left(\frac{r}{R_{inj}}
\right)^{3/2}\right] +\left (\frac{r}{R_{inj}}\right )^{1/2}\right ]^{-1}~.
\label{eq:dif_sol}
\end{equation}
In the absence of synchrotron energy losses, particle momenta 
only change due to adiabatic compression, and the momentum of a particle 
changes according with the well known $p=p_{inj}(r/R_{inj})^{-1/2}$, valid for
the case of free fall.

The joint effect of the energy gain due to the adiabatic compression and the 
energy losses due to synchrotron emission generates a new energy scale 
$p_m$ in the system:
$p_m=\frac{3 r c}{2 k_0 R_g^2}=\frac{3\pi r (m c)^2}{\sigma_T \dot{M}}$,
where $k_0=\sigma_T B_0^2/6\pi (mc)^2$ ($k_0=0$ no synchrotron losses).
Introducing $p_m$ we can rewrite $p[r,p_{inj},R_{inj}]$ as:
\begin{equation}
p[r,p_{inj},R_{inj}] =  p_{inj}
\left[\frac{p_{inj}}{p_m}\left(1-\left(\frac{r}{R_{inj}}\right)^{3/2}\right )
+ \left (\frac{r}{R_{inj}}\right )^{1/2}\right]^{-1} \ .
\end{equation}

From this expression it is clear that, at any fixed position $r$, 
adiabatic compression dominates over synchrotron losses if the injection
momentum is lower than $p_m$. In this case, the electron energy increases
while the electron moves inward, until the rate of synchrotron losses
become important. The opposite happens when the electrons are injected
at momenta larger than $p_m$, since synchrotron losses are important
from the time of injection.
The momentum $p_m$ can be interpreted as the momentum where the two 
competitive effects of adiabatic heating and synchrotron losses balance 
each other. Thus, particles accumulate at momentum $p_m$. This phenomenon 
depends on the distance from the galactic center: at large distances from 
the BH the momentum $p_m$, which scales linearly with radius, is large 
and the local rate of injected electrons is low, therefore, the accumulation 
is small. At small distances the accumulations at  $p_m$ grows.

We can define the electron equilibrium spectrum $n(E,r)$ which is related 
to $f(E,r)$ through the relation
$n(E,r)dE=4\pi p^2 f(p,r)dp$. 
In Fig. \ref{fig:nEr} we plot $n(E,r)$, in the two cases of an NFW (right 
panel) and a spike (left panel) density profiles, as a function of the 
electron energy at two different radii, $r=10^{3}R_g$ (upper solid curve) 
and $r=10^{4}R_g$ (lower solid curve); we have fixed $m_\chi=100$ GeV and 
$\langle \sigma v \rangle_{Ann}=10^{-27}$cm$^3$/s. 
The dashed lines illustrate the 
solution of the transport equation when adiabatic compression is switched off 
and only synchrotron losses are included. The accumulation effect described 
above manifests itself through the appearance of the spiky structure at 
momentum $p_m$. The accumulation is less pronounced at large radii, 
as expected.

We conclude this section by addressing the issue of the synchrotron self-
Compton scattering. As explained above, this effect cannot be accounted for
in an analytical approach to the transport equation, since it is intrinsically
nonlinear. Instead, we check a posteriori if neglecting SSC was a good 
assumption. The photon energy density as a function of $r$, 
$U_{ph}^{syn}(r)$, is easily calculated on the basis of symmetry arguments:
\begin{equation}
U_{ph}^{syn}(r)=\frac{1}{c}\left [ 
\frac{1}{r^2}\int_{r_{min}}^r dr' r'^2 \int d\nu j(\nu,r') 
+\int_r^{\infty}dr' \int d\nu j(\nu,r') \right ]~,
\label{eq:Uph}
\end{equation}
where $j(\nu,r)$ is the synchrotron emissivity. 

\begin{figure}[t!]
\epsfxsize=8cm   
\epsfbox{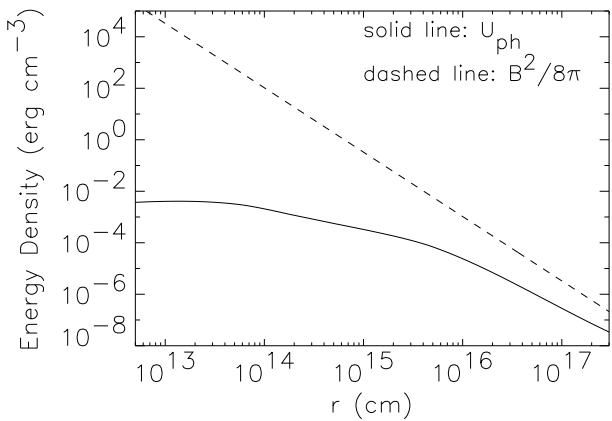}
\caption{Comparison between the energy density of the synchrotron emitted
photons $U_{ph}$ (solid line) and the energy density associated to the 
magnetic field $B^2/8\pi$ (dashed line).}
\label{fig:Uph}
\end{figure}

This energy density can now be compared with the magnetic energy density 
at the same location, $B^2(r)/8\pi$.
The results are plotted in Fig. \ref{fig:Uph}: the solid
line represents the photon energy density $U_{ph}^{syn}(r)$, while the magnetic
energy density is plotted as a dashed line. The calculations are
carried out for a dark matter density profile with the spike at the center
and a neutralino mass of 100 GeV. The magnetic energy always dominates over
the photon energy, although at large distances from the BH the difference 
between the two curves reduces to about one order of magnitude. The curves 
in Fig.\ref{fig:Uph} are obtained without taking into account the synchrotron
self-absorption effect, therefore the calculated
photon energy density should be considered as an upper limit. Thus, neglecting 
SSC is a good approximation for the present scenario.

Let us conclude this section addressing the issue of the synchrotron 
self-absorption (SSA) mechanism. Synchrotron radiation can be reabsorbed by 
the radiating electrons when the system is sufficiently compact. 
This phenomenon is particularly effective at low frequencies. 

Following the 
standard procedure \cite{ryb} we have included the SSA effect in the emission 
evaluation finding that this effect is efficient only in the case of the spike 
density profile and only for frequencies $\nu<10^{11}$ Hz \cite{noi2}. 
This results holds only in the case in which both synchrotron losses 
and adiabatic compression are taken into account. On the other hand if 
only synchrotron losses are considered the SSA effect becomes important 
even at lower frequencies. The dependence of the SSA on the neutralino 
parameters scales with 
$n(E,r)\propto \langle \sigma v \rangle_{Ann} m_{\chi}^{-3/2}$. 
At fixed $\langle \sigma v \rangle_{Ann}$, the electron density drops 
with increased $m_{\chi}$, and the SSA effect decreases as well. Finally, 
in the case of an NFW density profile without a spike, the rate of electrons 
injection remains such that the effect of SSA can always be neglected 
at the frequencies of interest.

\section{Synchrotron Emission from the GC}

In this section, we present our results in terms of the synchrotron luminosity 
produced by relativistic electrons from neutralino annihilations in the GC. 
We have considered both cases of a spike density profile and an NFW density 
profile. Taking into account the effect of SSA, that is efficient only 
in the spike case as discussed before, we have computed, through the 
equilibrium spectrum of electrons $n(E,r)$, the synchrotron luminosity
$L_{\nu}$ integrated over the all sky (i.e. over all lines of sight).  

\begin{figure}[t!]
\begin{tabular}{ll}
\epsfxsize=6cm   
\epsfbox{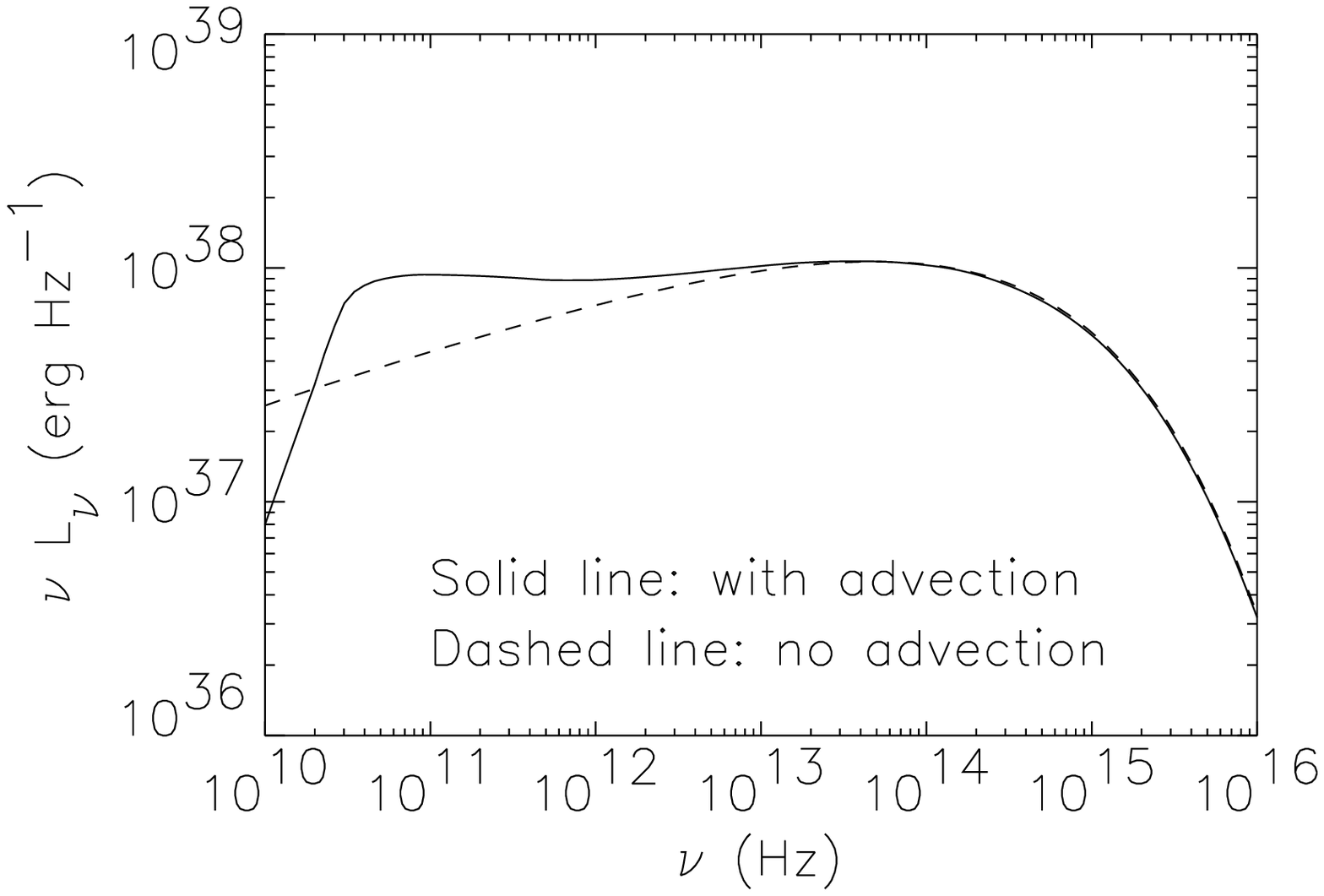} &
\epsfxsize=6cm   
\epsfbox{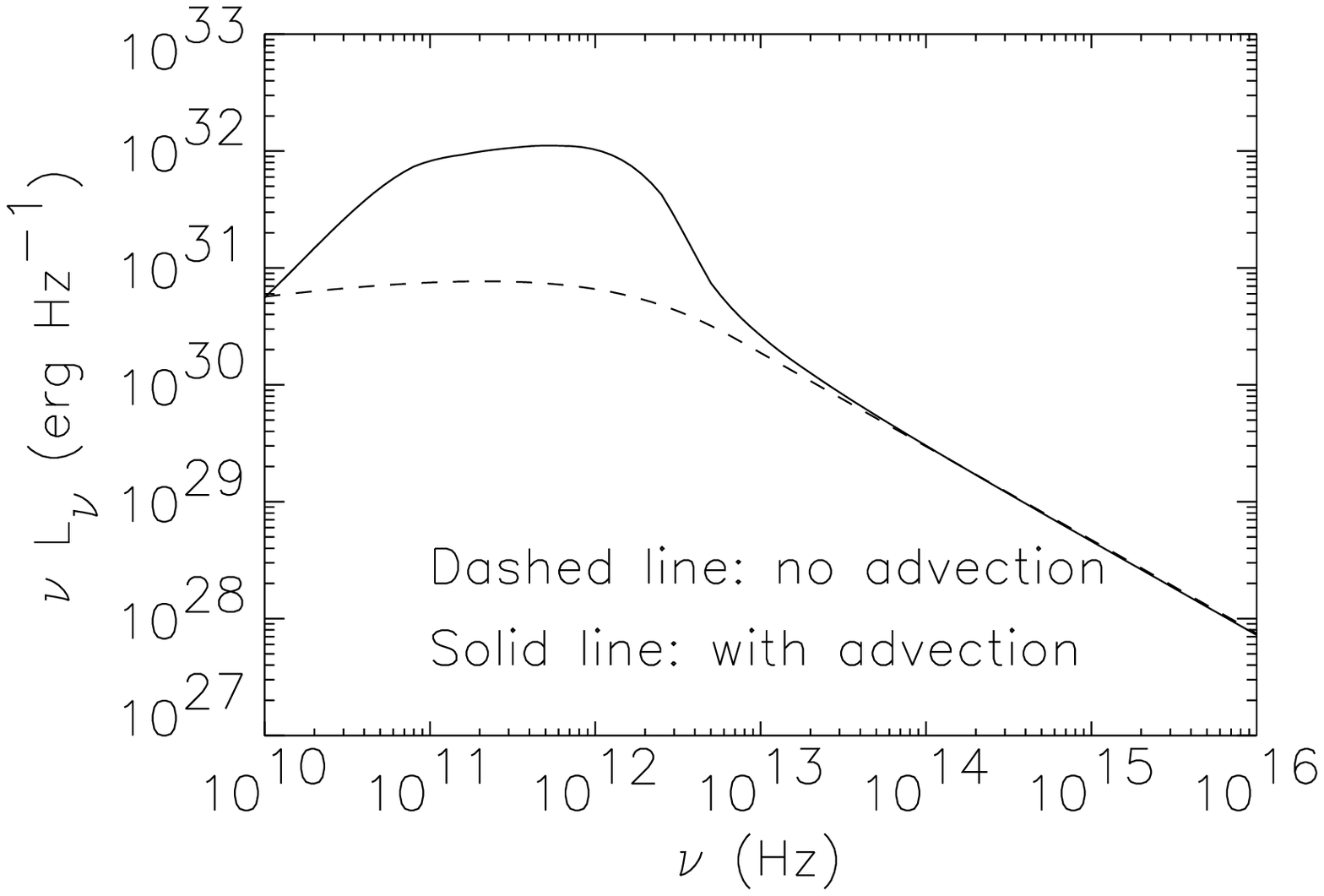} \\
\end{tabular}
\caption{
Emitted luminosty in two cases: 1) advection and synchrotron losses 
(solid line); 2) only synchrotron losses (dashed line). The computation is 
performed with $m_{\chi}=10^2$ GeV and 
$\langle \sigma v \rangle_{Ann}=10^{-27}$ cm$^3$/s. The left panel 
represents the Spike case while right panel the NFW case.}
\label{fig:Lnu1}
\end{figure}

In Fig. \ref{fig:Lnu1}, we fix  $m_{\chi}=10^2$ GeV and 
$\langle \sigma v \rangle_{Ann}=10^{-27}$ cm$^3$/s and show  the luminosity 
obtained with both the spike (left panel) and NFW (right panel) density 
profile for two cases: the continuous line shows the case in which advection 
is taken into account, while the dashed line shows the case in which only 
the synchrotron energy losses are present. 
Comparison of the two curves in Fig. \ref{fig:Lnu1} shows that the resonant 
behavior due to the combined effect of advection and synchrotron losses 
becomes important for frequencies up to $10^{14}$ Hz. In this frequency 
range, the electron density is much higher when advection is included as 
compared to the pure synchrotron case. This effect produces an increase in 
the emitted luminosity of about one order of magnitude. In addition, we can 
see that SSA decreases the emitted radiation in the frequency range 
$10^{10} - 10^{11}$ Hz. Moreover, the SSA effect becomes relevant only when 
advection is taken into account and only in the case of the spike density 
profile.

\begin{figure}[t!]
\begin{tabular}{ll}
\epsfxsize=6cm   
\epsfbox{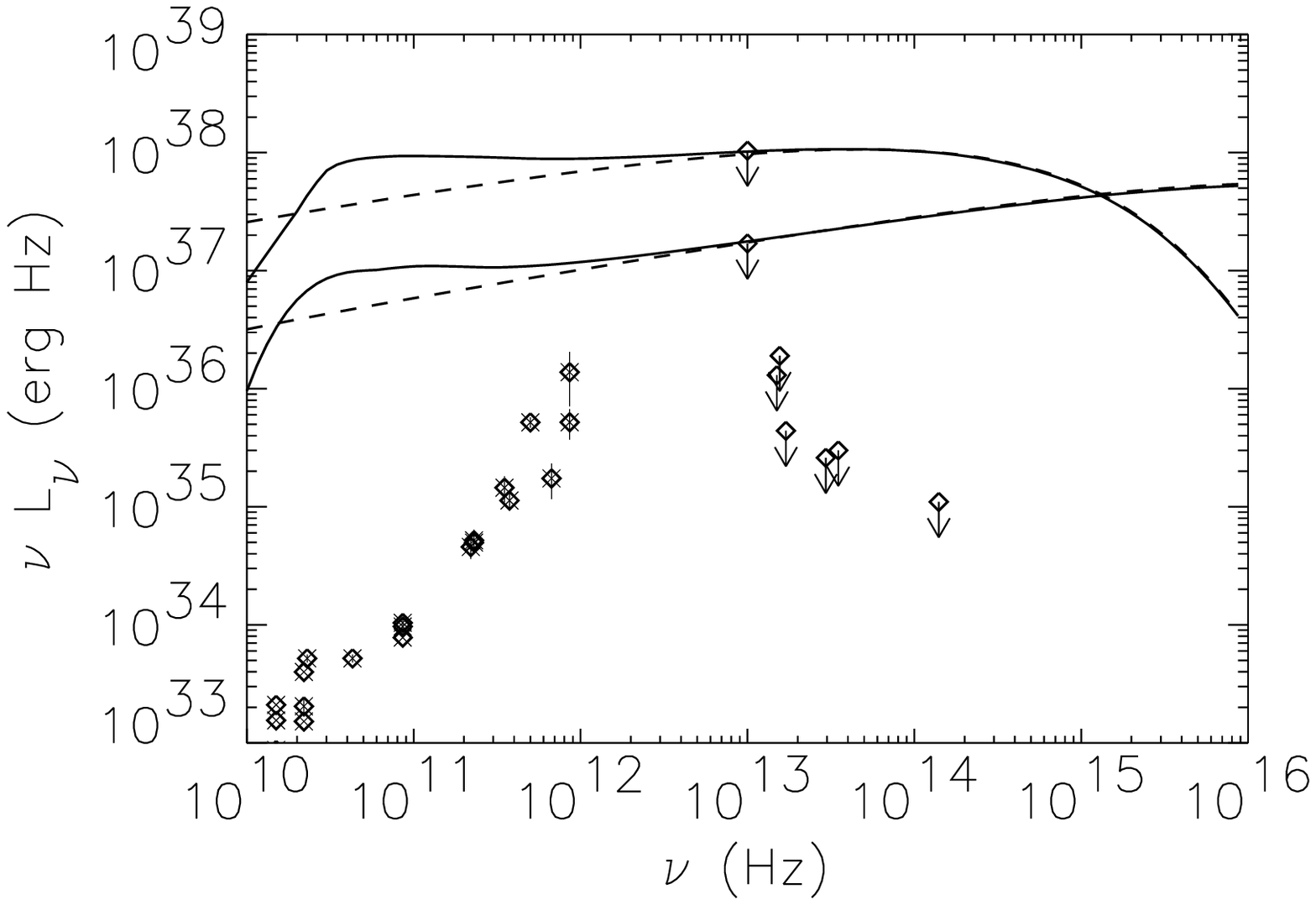} &
\epsfxsize=6cm   
\epsfbox{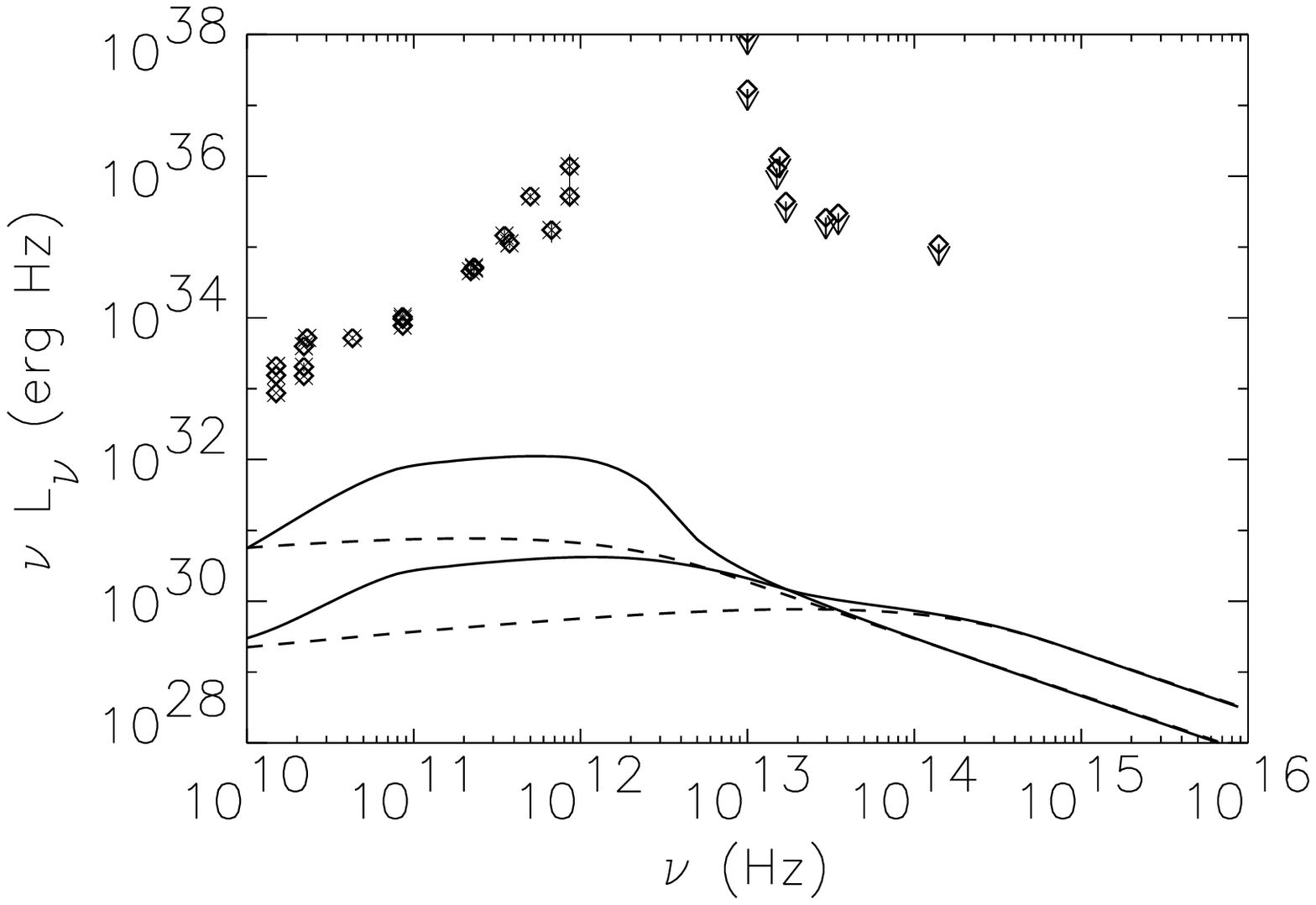} \\
\end{tabular}
\caption{
Luminosity, compared with experimental data from Sgr A$^*$,
in the case of the spiky density profile (left panel) and NFW density
profile (right panel). Dashed curves represent the luminosity obtained
neglecting the effect of advection. The computation is performed with 
$m_{\chi}=10^2$ GeV (upper curves) and $m_{\chi}=10^3$ GeV (lower curves). 
The annihilation cross section is 
$\langle \sigma v \rangle_{Ann}=10^{-27}$ cm$^3$/s.} 
\label{fig:Lnu2}
\end{figure}

In order to constrain the shape of the dark matter density profile at 
the GC or the parameters of the dark matter particle, we compare the 
calculated luminosities with the observations. The well-studied source at 
the GC, Sgr A$^*$, has been observed in a large range of frequencies, from 
the radio up to the near-infrared. Experimental data from $10^9$ Hz to 
$10^{14}$ Hz \cite{data} are displayed in Fig. \ref{fig:Lnu2}. In the same 
figure we show the two cases of spiky (left panel) and NFW (right panel) 
density profile, for $m_{\chi}=10^2$ GeV (upper curves) and 
$m_{\chi}=10^3$ GeV (lower curves) and 
$\langle \sigma v \rangle_{Ann}=10^{-27}$ cm$^3$/s. 
In Fig. \ref{fig:Lnu2} we also show the results for the cases in which only 
synchrotron losses are taken into account (dashed curves).
From Fig. \ref{fig:Lnu2}, it is clear that a spike in the GC induces much 
stronger emission than the observed flux. Therefore, either there is no 
spike in the dark matter profile or neutralinos are not the dark matter. 
This conclusion agrees with previous studies \cite{gon2} that show how 
changes to the neutralino parameter do not gap this large discrepancy. 
The parameter space for neutralinos 
$m_{\chi},\langle \sigma v \rangle_{Ann}$
has been studied extensively \cite{gon2}. From accelerator studies 
$m_{\chi}=10^2$ GeV is a lower limit and from cosmological constraints 
(i.e., dark matter density) $m_{\chi}$ and $\langle \sigma v \rangle_{Ann}$  
must lie in the range: $m_{\chi}\in [10^2 {\rm GeV},10^3 {\rm GeV}]$ and
$\langle \sigma v \rangle_{Ann} \in [10^{-27} {\rm cm}^3{\rm /s}, 10^{-26} 
{\rm cm}^3{\rm /s}]$ \cite{Baltz}. 

The emitted luminosity scales with the 
neutralino mass and the annihilation cross section through the electron 
density $n(E,r)$, following the same relations obtained in Eq. 
(\ref{scalings}). This implies that the luminosity can only be lowered 
by at most one order of magnitude, by increasing the neutralino mass up to 
$\sim$ 1 TeV. While it is easy to increase the luminosity by an order 
of magnitude without violating any bound simply by increasing the cross 
section of neutralino annihilation, it is hard to decrease it. In fact, 
as $\langle \sigma v \rangle_{Ann}$ is lowered well below 
$10^{-27}$ cm$^3$/s, the density of neutralinos becomes smaller than the 
necessary density to account for cold dark matter.

The large enhancement of the neutralino density due to the spike profile 
produces a synchrotron luminosity that is difficult to reconcile with 
observations. Moreover, this conclusion only gets stronger when advection is 
included. The SSA effect does not affect the discrepancy between a neutralino 
spike and the observations, since it changes the luminosity only at 
frequencies in the range $10^{10}\div 10^{11}$ Hz.
The situation is completely different for the case of the NFW density profile. 
In this case, the synchrotron luminosity is always less than the experimental 
data as can be seen in Fig. \ref{fig:Lnu2}. NFW is consistent with 
observations even if  $\langle \sigma v \rangle_{Ann}=10^{-26}$ cm$^3$/s 
is considered.

In this paper, we only considered two possibilities for the dark matter 
density profile: the less concentrated hypothesis (NFW) and the most 
concentrated one (spike). Other proposed profiles such as the Moore et al. 
profile should generate a luminosity in between the spike and the NFW cases. 
In this case, observations are likely to place more stringent limits on the 
neutralino parameters instead on the density profile itself. 

\section{Conclusions}

In this paper we have discussed two possible avenues that lead to  
an indirect detection of the annihilation of neutralino CDM. We have
discussed the effect of this annihilation in the halo of our galaxy
as well as in its center. 

The expected emission from the halo annihilation drops into the $\gamma$-ray
frequency range and it generates a diffuse $\gamma$-ray background. The Dark 
matter sub-structures of the galactic halo can be a dominant component of the 
diffuse $\gamma$-ray background depending on the concentration and location
of DM clumps in the inner regions of the galactic halo. In order to bracket
the range of possible fluxes we have considered two extreme scenarios, that 
we named type I and type II. The first corresponds
to  extremely concentrated clumps present everywhere in the Galaxy halo,
while in the second scenario the clumps are much less concentrated and are
completely destroyed by tidal effects in the inner 10 kpc of the Galaxy. While
the type I scenario allows one to put very strong constraints on the
properties of neutralinos and on the density profile inside clumps, the type
II scenario generates  fluxes of diffuse $\gamma$-rays which are  barely
detectable, with the exception  of the case in which the density profile of
the clumps is modeled as a SIS sphere. For the type I scenario, most of the 
parameter space of neutralino dark matter is ruled out if the density profile 
of dark matter clumps is in the form of a SIS sphere or a Moore profile. 
Weaker bounds can be imposed on the neutralino parameter space in the case 
of NFW density profile.

The second possibility that we have considered here is related to the 
electron-positron emission in the GC magnetic field. Electron-positron pairs
are expected to be copiously produced in neutralino annihilation, from the 
comparative study of their synchrotron emission at the GC with the 
experimental data, already available in the radio and near IR frequency 
range, we have obtained severe limits on the neutralino density profile at the 
GC. Using our results we can reaffirm that a spike density profile is ruled 
out by these radio and near IR observations. We reached this conclusion by 
a careful consideration of the accretion flow and the loss processes in the 
transport equations for the neutralino generated electrons and positrons as 
well as the radiative transfer. However, rejecting the spike hypothesis the 
observed emission from the GC is consistent with neutralinos following an NFW
density profile. We can conclude stating that other proposed density profiles,
like the Moore et al., should generate an emission in between the two profiles
we have considered here. In this case the experimental data from Sg A$^*$ 
will put severe constraints on the neutralino parameters. 

\section*{Acknowledgments}
I would like to express my special thanks to Pasquale Blasi and 
Angela V. Olinto with whom the present work was developed.

\end{document}